%Paper: hep-th/9308117
%From: MANSFIELD%HEP.DURHAM.AC.UK@ib.rl.ac.uk
%Date: Wed, 25 Aug 93  17:42 BST
%Date (revised): Sat, 4 Sep 93 20:19 BST

\def\newhat #1{\mathaccent 94#1}
\def\D{{\cal D}}
\def\met{g_{ab}}
\magnification=\magstep 1
\null\vskip 0.75in
{\hfil DTP-93-45}
\bigskip
\centerline{\bf The Consistency of Topological Expansions in Field Theory:}
\centerline{\bf `BRST Anomalies' in Strings and Yang-Mills}
\vskip 0.5in
\centerline{Paul Mansfield}
\bigskip
\centerline{Department of Mathematical Sciences,}
\centerline{University of  Durham,}
\centerline{South Road,}
\centerline{Durham, DH1 3LE}

\vskip 0.5in

\centerline{\bf Abstract}
\bigskip
\noindent
Many field theories of physical interest have configuration spaces
consisting of disconnected components. Quantum mechanical amplitudes are
then expressed as sums over these components. We use the Faddeev-Popov
approach to write the terms in this topological expansion as moduli space
integrals. A cut-off is needed when these integrals diverge. This
introduces a dependence on the choice of parametrisation of configuration
space which must be removed if the theory is to make physical sense. For
theories that have a local symmetry this also leads to a breakdown in BRST
invariance. We discuss in detail the cases of Bosonic Strings and
Yang-Mills theory, showing how this arbitrariness may be removed by the
use of a counter-term in the former case, and by compactification on $S\sp 4$
in the latter.
\vfill
\eject

\centerline{\bf The Consistency of Topological Expansions in Field Theory}
\centerline{\bf `BRST Anomalies' in Strings and Yang-Mills}

\bigskip
\centerline{\bf 1. Introduction}
\bigskip
   Despite their very different physical interpretations there are many
similarities between the mathematical structures of Yang-Mills theory and
first quantised String Theories. Both have local symmetries which  are
crucial to their consistency.
For Yang-Mills theory this is gauge invariance, for String Theory this is
invariance under world-sheet reparametrisations.
After gauge-fixing and the introduction of
Faddeev-Popov [1] ghosts they are BRST invariant [2],
and this may be made the basis of a quantisation procedure.
Both theories are invariant classically under local scalings of the metric.
In Yang-Mills theory this is useful in constructing classical solutions
and the breaking of the invariance at a quantum level underlies the use of
the renormalisation group. For critical strings the maintenance of invariance
under
Weyl transformations of the world-sheet metric is a crucial constraint on
the quantisation of the system [3].
Also both theories have topologically non-trivial sectors so that any
transition amplitude is an infinite sum over contributions from these
sectors weighted with an appropriate coupling constant. For closed strings
this is a sum over closed Riemann surfaces of increasing genus
corresponding to loops of virtual strings [4]. The coupling constant is
the  string coupling, $\kappa$, to the power of minus the Euler
characteristic which counts the number of three-string-interactions
vertices.
In the Euclidean formulation of Yang-Mills theory stereographically
projected onto $S\sp 4$ the configurations of the gauge-potential fall into
distinct homotopy classes [5]. These
are classified by the second Chern class or instanton number, and the
coupling is essentially the theta angle. The coupling constant dependence
of these two expansions appears to result from very different physical
considerations but in both cases it may be ultimately  traced back to
unitarity.  Furthermore the functional integral for a fixed
topological sector may be reduced to a finite dimensional integral over a
number of parameters or moduli. In String Theory these are the moduli of
the Riemann surfaces, in Yang-Mills theory they are the instanton moduli.
Typically these finite dimensional integrals diverge in some region of the
moduli space, unless there are good reasons otherwise, for example
supersymmetry in the case of strings.
The one instanton sector contribution to the partition function of Yang-Mills
theory for the gauge group $SU(N)$
reduces to an integral over the instanton position $y\sp \mu$ and scale
$\rho$. When space-time is taken to be $R\sp 4$ this integral is, in the
semi-classical approximation at one loop [6],

$$Z_1\sp {YM}=e\sp {-{8\pi\sp 2\over g\sp 2(\mu )}}\int
{d\sp 4y\, d\rho\over\rho\sp 5}\rho\sp {11N/3}.\eqno (1.1)$$

This diverges for large scales. The integral over position is less
troubling since it may reasonably
be taken to yield the volume of space-time, in which case $Z_1$ is the
integral of a constant density. Both divergences are infra-red, or large
distance effects and for this reason are not considered as pathologies
of the theory so much as defects of the approximation, which is
only considered valid at short distances where asymptotic freedom holds
sway. The one string loop partition function is an integral over the
complex modulus $\tau$ [7]

$$Z_1\sp {string}=\int_F d\sp 2\tau\, (\Im \tau )\sp {-2}C(\tau ),\quad C(\tau)
=4({1\over 2}\Im \tau )\sp {-12}e\sp {4\pi\Im\tau  }|\prod_{n=1}\sp \infty
(1-e\sp {2n\pi i\tau})|\sp {-48}. \eqno (1.2)$$
The
fundamental domain is usually taken to be $F:\quad -{1\over 2}\le \Re
\tau\le{1\over 2},\quad\Im\tau >0,\quad |\tau |\ge 1$. The integral
diverges as $\Im \tau $ becomes large. This is again interpreted as an
infra-red divergence, although by modular invariance it may be mapped to a
small $\tau$ one if the integral is taken over a different, but
equivalent, domain.
These divergences are the subject of
this paper.
\medskip
             Given an (ill-defined) integral over a domain $M\sp \prime$,
$\int_{M\sp \prime}d\sp nt\, f(t)$, that diverges in some region the simplest
thing to do is to introduce a cut-off by restricting the range of values
of the variables $t$. Suppose we do this by restricting just one of the
variables, $t\sp 1$ say. Then $M\sp \prime$ is replaced by a new domain $M$
with
a boundary $\partial M$ on which $t_1$ takes its cut-off value. The
well-defined integral $\int_M d\sp nt\, f(t)$ now depends strongly on the
value of the cut-off and our choice of the parameter $t\sp 1$. If we were to
make a reparametrisation $t\sp A\rightarrow \tilde t\sp A=t\sp A+\epsilon\sp A
(t)$
then the cut-off changes and consequently so does the value of the
integral.
The change in the integral can be expressed using Stokes' theorem as

$$\int_M d\sp nt\,{\partial \over\partial t\sp A}\left(\epsilon \sp A(t)\,
f(t)\right)=
\int_{\partial M}d\Sigma_A\,\epsilon\sp A (t)\, f(t).\eqno (1.3)$$
If $\int d\sp nt\, f$ is the contribution to an amplitude from a particular
topological sector then we have a regulated expression that depends on our
choice of parametrisation if (1.3) is non-zero. This is not acceptable and
a way must be found of removing this parametrisation dependence from the
theory. Now the field configurations responsible for the divergences are
those that degenerate to configurations belonging to a different
topological sector. In the String Theory case we will see that the one
string loop contribution to scattering amplitudes is proportional to the
tree-level, or zero-loop, contribution as $\Im \tau\rightarrow\infty$, and in
the case of Yang-Mills theory the one-instanton solution to the classical
equations of motion approaches the (vanishing) zero-instanton solution as
$\rho\rightarrow\infty$. Care must be exercised in making these statements
precise because the degenerating configurations belong to the same
topological class, characterised by the same invariants, as the
non-degenerate configurations, but roughly speaking the field
configurations corresponding to $\partial M$ approximate to configurations
belonging to a different topological class. This means that it may be
possible to cancel (1.3) with a counter-term added to the contribution
from a different topological sector, leading to a renormalisation of the
topological expansion parameter. This `topological renormalisation'
can be implemented for the Bosonic String, although we will see that it is
unnecessary for Yang-Mills theory. There is a number of other field
theories that possess the same topological characteristics, for example
Higgs Models and $CP\sp n$-models, so in the next section we will present a
general formulation of the reduction of the path-integral for a
topologically non-trivial sector to an integral over moduli and discuss
its parametrisation dependence.

\vfill
\eject
\centerline{\bf 2.  General Formulation}
\bigskip
  The general setting in which we are interested is a theory with fields
$\phi$ and action $S[\phi ]$ for which amplitudes are expressed as sums of
functional integrals each of which is over a distinct homotopy class of
configurations of $\phi$. Suppose these classes, ${\cal C}_n$, are
labelled by an integer $n$, then if we concentrate to begin with on the
partition function $Z$

$$Z=\sum_n\, \kappa_n\,Z_n,\quad Z_n=\int_{{\cal C}_n}{\cal D}\phi\,
e\sp {-S[\phi ]}\eqno (2.4).$$
Here
$\kappa_n$ is some function of the topological coupling constant. In practice
we usually compute functional integrals by semi-classical expansions.
Within each sector ${\cal C}_n$ there will be a solution,
$\phi_0$, to the classical
equations of motion depending, in general, on a finite number of parameters
$\{ t\sp A \}$, so we set $\phi=\phi_0(t)+\bar\phi$. $\bar\phi$ is
continuously deformable to zero, so we could expand the action in powers
of $\bar\phi$ to obtain a Gaussian to leading order, the linear term
vanishing by the equations of motion. However the derivatives of $\phi_0$
with respect to the moduli will be zero-modes of the quadratic action, so
that $\bar\phi$ should be made transverse to them by imposing some
constraints. The integral over $\bar\phi$ can then be done yielding a
function of the $t$. This must then be integrated over the moduli
corresponding to the remaining degrees of freedom of $\phi$, the
zero-modes. We want to liberate ourselves from the restrictions of the
semi-classical expansion so we will extract this dependence on the moduli
using the Faddeeev-Popov trick without resorting to approximation. This is
particularly useful because in the two cases in which we are primarily
interested there is also a gauge-invariance that can be treated
simultaneously using this method. So we will suppose that the action has
an infinite number of invariances parametrised by group elements $g$
i.e. if $\phi\rightarrow\phi\sp g$ then $S[\phi ]\rightarrow S[\phi\sp
g]=S[\phi
]$. These symmetries will form a closed algebra, so for $g$ close to the
identity, $g=1+\omega$, and $\omega\sp a$ the components of $\omega$ in some
basis then $\delta_\omega\phi = \omega\sp a\delta_a\phi$ and

$$[\delta_a,\delta_b]\phi=
f_{ab}\sp c\delta_c\phi\eqno (2.5)$$
We
will introduce the moduli and gauge-fixing conditions simultaneously.
Choose an infinite number of functions $F_j(\phi ,t)$ such that the
conditions $F_j(\phi\sp g,t)=0$ for any
$\phi\in {\cal C}_n$  have (locally) a unique solution for
$g$ and the moduli ${t\sp A}$.
(In general there will be a Gribov ambiguity
globally.) For example in the Yang-Mills case we could choose the
conditions to consist of  the background gauge condition as well as a
finite number of constraints requiring $\bar\phi$ to be orthogonal to the
zero-modes. With ${\bf A}_\mu$ denoting the gauge-potential, and
${\cal A}_\mu (t)$
the instanton solution to the classical equations of motion depending on
moduli $t$, we would take on $R\sp 4$

$$[\partial_\mu +{\cal A}_\mu (t), {\bf A}_\mu -{\cal A}_\mu (t)]=0 ,
\quad \int d\sp 4x\,tr\,\left({\partial {\cal A}_\mu (t)\over\partial t\sp A}
\left(
{\bf A}_\mu-{\cal A}_\mu (t)\right)\right)=0.\eqno (2.6)$$
Following
the usual Faddeev-Popov construction [1] we
want to impose these constraints using delta-functions, so to this end we
look for a functional $\Delta [\phi, t]$ such that

$$\int{\cal D}g \, dt\,
\Delta [\phi ,t ]\prod_j\delta
(F_j(\phi\sp g,t))=1.
\eqno (2.7)$$
Here ${\cal D}g$ is the Haar measure on the group of gauge tranformations.
Invariance of this measure under multiplication of $g$ by a group element
implies that $\Delta [\phi\sp g,t]=\Delta [\phi
,t]$.
Suppose that for $\phi=\tilde\phi$ the constraints have a solution
$g=\newhat  g$ and $t=\newhat  t$.
If
we expand about this solution $g=(1+\omega )
\newhat  g$ and $t=\newhat  t +\tilde t$ then, with $\partial_A$ denoting
$\partial/\partial t\sp A$

$$F_j(\tilde \phi\sp g ,t)\simeq \left(\delta_\omega +\tilde t\sp A{\partial_A}
\right) F_j(\phi ,t )|_{\phi=\tilde\phi\sp {\newhat  g},t=\newhat  t}.\eqno
(2.8)$$
so
that

$$\int {\cal D}g\, dt\, \prod_j\delta\left(F_j(\tilde\phi\sp g,t)\right)=
\int {\cal D}\omega\, d\tilde t\,\prod_j\delta \left((\delta_\omega+
\tilde t\sp A{\partial_A})F(\phi ,t)|_
{\phi=\tilde\phi\sp {\newhat  g},t=\newhat  t}\right)\eqno (2.9)$$
By the usual rules for
integrating out delta-functions
(2.9) becomes

$$Det\sp {-1}\left( \delta_a F_j (\phi ,t) ,\quad {\partial_A} F_j(\phi ,t)
\right)|_{\phi=\tilde\phi\sp {\newhat  g},t=\newhat  t},\eqno (2.10)$$
so that
$\Delta$ is the determinant itself. This may be represented using
Grassmann numbers. For each transformation parameter $\omega\sp a$ we
have a ghost $c\sp a$, for each constraint $F_j$ an anti-ghost $b\sp j$ and for
each
modulus $t\sp A$ a quasi-ghost $\tau\sp A$, enabling the determinant to be
represented as

$$\Delta =\int {\cal D}(b,c)\,d\tau\,exp\,
-\left((c\sp a\delta_a +\tau\sp A{\partial_A}) (b\sp j\,F_j)\right).
\eqno (2.11)$$
Inserting
the identity (2.7) into $Z_n$ and changing the order of integration gives

$$Z_n=\int{\cal D}g\, dt\int{\cal D}\phi\,e\sp {-S[\phi]}\Delta [\phi  ,t]
\prod_j\delta\left( F_j(\phi\sp g, t)\right)\eqno (2.12)$$
If we take $\phi\sp g$ as a new integration variable then assuming that
${\cal D}\phi={\cal D}\phi\sp g$ and using the invariance of $\Delta$ and
$S[\phi]$ we obtain, on renaming $\phi\sp g$ as $\phi$

$$Z_n=\int{\cal D}g\, dt\int{\cal D}\phi\,e\sp {-S[\phi]}
\Delta [\phi  ,t]
\prod_j\delta\left( F_j(\phi, t)\right)
\equiv  \left(\int{\cal D}g\right) \int dt\,z(t)\eqno (2.13)$$
where
the infinite volume of the gauge group is now explicitly factored out.
With the representation (2.11)
and writing the delta-functions as integrals over $\lambda\sp j$
we finally arrive at the moduli space
density $z(t)$ expressed as

$$z(t)=\int {\cal D}(\phi , b,c,\lambda )\, d\tau
\,e\sp {-S_{FP}}$$

$$S_{FP}\equiv
S[\phi ]+i\lambda\sp j F_j (\phi ,t )+(c\sp a\delta_a+\tau\sp A\partial_A )
b\sp j\,F_j(\phi ,t ) ,\eqno (2.14)$$
This
can be written more economically using the BRST transformation [2]. In our
case, due to the presence of the moduli, the BRST transformation will {\it
not}
be a symmetry of $S_{FP}$, although it is still useful. This
transformation is parametrised by a Grassmann number, $\eta$ say, and acts
on the fields $\phi,b,c,\lambda,\tau$, but not on the moduli, and may be
written as $\delta_\eta\phi=\eta\varsigma\phi$ etc., where the $\varsigma$
operation is given by

$$\varsigma\phi=c\sp a\delta_a\phi,\quad\varsigma c\sp a={1\over
2}c\sp bc\sp cf\sp a_{bc},\quad\varsigma
b\sp j=i\lambda\sp j,\quad\varsigma\lambda\sp j=0,\quad\varsigma\tau\sp A=0.
\eqno (2.15)$$
$\varsigma$ is nilpotent by construction, i.e. $\varsigma\sp 2=0$.
With its use we can write the action as

$$S_{FP}=S[\phi ]+(\varsigma +\tau\sp A\partial_A)(b\sp j\,F_j(\phi ,t)).
\eqno (2.16)$$
$\varsigma$ and $\tau\sp A\partial_A$ anti-commute
because $\partial_A$ acts only on the moduli whereas $\varsigma$ does not.
Furthermore $\tau\sp A\partial_A$ is nilpotent since derivatives with
respect to the moduli commute with each other, so the sum $\varsigma+
\tau\sp A\partial_A$ is also nilpotent. Given that $S[\phi ]$ is gauge
invariant and independent of the moduli, which only enter $S_{FP}$ via the
constraints, it follows that

$$(\varsigma +\tau\sp A\partial_A)\,S_{FP}
=(\varsigma +\tau\sp A\partial_A)\,S[\phi]+
(\varsigma +\tau\sp A\partial_A)\sp 2\,b\sp j\,F_j(\phi ,t )=0\eqno (2.17)$$
so that
the gauge-fixed action is not BRST invariant, but rather
$\varsigma S_{FP}=-\tau\sp A\partial_A S_{FP}$. If we denote the fields
$\phi,b,c,\lambda,\tau$ collectively by $\Psi$ then the Jacobian for the
transformation $\Psi\rightarrow\Psi+\delta_\eta\Psi$ is
one plus the super-trace of the derivative of $\delta_\eta\Psi$ with
respect to $\Psi$. This super-trace is

$${\delta\over\delta\phi}\eta c\sp a\delta_a\phi -{\partial\over\partial
c\sp a}\left({\textstyle 1\over 2}\eta c\sp bc\sp cf\sp a_{bc}\right)=
\eta c\sp a\left({\delta\over\delta\phi}\delta_a\phi+f\sp b_{ba}\right)
\eqno (2.18)$$
so
the volume element ${\cal D}\Psi $ is invariant if
$${\delta\over\delta\phi}\delta_a\phi+f\sp b_{ba}=0.\eqno (2.19)$$

\medskip
     The partition function is given by a sum of moduli-space integrals
$\int dt\,z(t)$, with $z(t)=\int {\cal D}\Psi \,exp-S_{FP}$. If these
integrals diverge then we cut off the region of
integration to $M$ introducing additional
components to the boundary $\partial M$. We need to know if this procedure
depends on the choice of arbitrary conditions $F_j$. If it does, then this
is  unsatisfactory  and we have to find some way of repairing the damage.
Suppose we make an arbitrary variation of the constraints $F_j\rightarrow
F_j+\delta F_j$, then the change in the action is $\delta S_{FP}=
(\varsigma +\tau\sp A\partial_A )b\sp j\delta F_j$ so

$$\delta e\sp {-S_{FP}}=
-(\varsigma +\tau\sp A\partial_A )(b\sp j\delta F_j)\,e\sp {-S_{FP}}=
-(\varsigma +\tau\sp A\partial_A )\left(\,b\sp j\delta F_j
e\sp {-S_{FP}}\right).\eqno (2.20)$$
Thus
$$\delta z(t)=-\int{\cal D}\Psi (\varsigma +\tau\sp A\partial_A)\left(
b\sp j\,\delta F_j\, e\sp {-S_{FP}}\right) .\eqno (2.21)$$
Consider $\int{\cal D}\Psi\, b\sp j\delta F_j\,exp-S_{FP}$. This vanishes
because it is Grassmann odd. If we change the integration variables
$\Psi\rightarrow\Psi+\delta_\eta\Psi$ then the change in the integral is

$$\int{\cal D}\Psi\, \eta\varsigma (b\sp j\delta F_j\,e\sp {-S_{FP}}),\eqno
(2.22)$$
but the
value of the integral does not change under a change of integration
variable, so (2.22) vanishes for all $\eta$. Consequently

$$\delta z(t)=-\partial_A
\int{\cal D}\Psi \tau\sp Ab\sp j\,\delta F_j\, e\sp {-S_{FP}}
.\eqno (2.23)$$
Using
Stokes' theorem the change in the density integrated over the cut-off
moduli space is

$$\delta\int_Mdt\,z(t)=-\int_{\partial M}d\Sigma_A
\int{\cal D}\Psi \tau\sp Ab\sp j\,\delta F_j\, e\sp {-S_{FP}}.\eqno (2.24)$$
If
this is not zero then we have a problem because the partition function
depends on our choice of constraints. On $\partial M$ at least one of the
moduli takes its cut-off value so the field configurations contributing to
$z(t)|_{\partial M}$ approximate to those of a different topological
sector, thus it may be possible to cancel (2.24) with a counter-term added
into the contribution to the partition function from this other sector.
The overall value of the partition function may then be made independent
of the choice of constraints. The details of how this is done will vary
from theory to theory. In section 3 we will see how the breakdown of BRST
invariance in the bosonic string at one-string-loop can be corrected for
by a tree-level counter-term.
\medskip
     More generally we should consider the contribution from ${\cal C}_n$ to
the expectation value of some
gauge-invariant operator $V(\phi )$,

$$\langle V\rangle _n\equiv\int dt\int {\cal D}\Psi \,V(\phi )\,e\sp {-S_{FP}},
\quad \delta_a V(\phi )=0\eqno (2.25)$$
Since
$V$ is gauge-invariant and independent of the moduli which only enter via
the constraints it is
annihilated by $\varsigma +\tau\sp A\partial_A$. Repeating the above argument
shows that the change in $\langle V\rangle_n$ resulting from a change in
the constraints is

$$\delta\langle V\rangle_n=
-\int_{\partial M}d\Sigma_A
\int{\cal D}\Psi\, V(\Psi) \tau\sp Ab\sp j\,\delta F_j\, e\sp {-S_{FP}}.\eqno
(2.26)$$
We
need to ensure that any counter-term constructed to re-instate
reparametrisation invariance in the sum over topological sectors for the
partition function has the effect of making expectation values of
gauge-invariant operators reparametrisation invariant too.

\medskip
      Formula (2.24) also describes the effect on the partition function
of a transformation,
$\phi\rightarrow\phi+\delta\phi$,
that is an additional symmetry of $S[\phi ]$ that
commutes with the gauge invariance. For example Weyl invariance in
Yang-Mills theory. Under this transformation the change in the
Faddeev-Popov action is entirely due to the change in $F_j$, and assuming
that $[\varsigma ,\delta]=0$ (2.24) gives the resulting change in the
contribution of ${\cal C}_n$ to the partition function provided the
Jacobian is unity.
\medskip
      The usual Ward identities that express the BRST symmetry [2]
are modified by
the above considerations. As a consequence the theory will not be BRST
invariant unless the invariance under the choice of $F_j$ is repaired.
In the usual case the Ward identities state that the expectation value of
an operator of the form $\varsigma V(\Psi )$ vanishes. This is crucial to
the usual perturbative renormalisation of Yang-Mills theory [8], and to the
decoupling of spurious states in String Theory, and so clearly is a
feature of the theory that we do not want to spoil. However, we will see
that the contribution of the topological sector ${\cal C}_n$ to this
expectation value reduces to an integral over $\partial M$. Consider
the contribution of ${\cal C}_n$ to
$\int{\cal D}\Psi\,V(\Psi )e\sp {-S_{FP}}$, with $V$ Grassmann odd. Applying
the
change of integration variables $\Psi\rightarrow\Psi+\delta_\eta\Psi$ on
the assumption that (2.18) holds we conclude that
$\int{\cal D}\Psi\,\varsigma(V(\Psi )exp-S_{FP})$
vanishes. Writing this out, and using the fact that $\varsigma +\tau\sp A
\partial_A$ annihilates $exp-S_{FP}$ we obtain

$$0=\int{\cal D}\Psi \,(\varsigma V(\Psi ))e\sp {-S_{FP}}
-\int{\cal D}\Psi \,V(\Psi )\left(\varsigma e\sp {-S_{FP}}\right)$$
$$=\int{\cal D}\Psi \,(\varsigma V(\Psi ))e\sp {-S_{FP}}
+\int{\cal D}\Psi \,V(\Psi )\,\tau\sp A\partial_A\,e\sp {-S_{FP}} .\eqno
(2.27)$$
The
contribution of ${\cal C}_n$ to the expectation value of $\varsigma V(\Psi )$
is
thus

$$\langle\varsigma V(\Psi )\rangle_n=
\int_M dt\int{\cal D}\Psi \,(\varsigma V(\Psi ))e\sp {-S_{FP}}=   -
\int_{\partial M}d\Sigma_A\int{\cal D}\Psi \,\tau\sp A\,V(\Psi )\,e\sp
{-S_{FP}}.
\eqno
(2.28)$$
If this does not vanish then we can think of it as a `BRST anomaly'.
It needs to be cancelled by a counter-term if we are to have all the usual
nice properties guaranteed by BRST invariance.

\bigskip
\centerline{\bf 3. Bosonic Strings}
\bigskip
 In Polyakov's formulation of String Theory [3] the partition function
for closed strings is given by a sum of functional integrals over closed
Riemann surfaces of increasing genus, $h$, weighted by a power of the
coupling $\kappa$ [4]

$$Z=\sum_{h=0}\sp \infty \kappa\sp {2-2h}\int_h \D\met\,\D X\sp \mu\,e\sp
{-S[\met
,X\sp \mu]}.\eqno (3.1)$$
These
integrals are rather formal because they require the factoring out of an
infinite gauge group volume. The fields $\met$ and $X\sp \mu$ are functions
of the world-sheet coordinates $\xi\sp a$ and describe an intrinsic metric and
the coordinates of a surface embedded in D-dimensional space-time with
metric $\eta_{\mu\nu}$,
respectively. The action is [9]

$$S[\met ,X\sp \mu ]={1\over 2}\int d\sp 2\xi\sqrt g g\sp {ab}
{\partial X\sp \mu\over\partial\xi\sp a}
{\partial X\sp \nu\over\partial\xi\sp b}\eta_{\mu\nu}.\eqno (3.2)$$
The
volume elements are constructed from inner products on variations of the
fields

$$(\delta X,\delta X)=\int d\sp 2\xi \,\delta X\sp \mu \delta X\sp
\nu\eta_{\mu\nu},
\quad (\delta g,\delta g)=\int d\sp 2\xi \,\delta \met\delta g_{cd}g\sp
{ac}g\sp {bd}
\eqno (3.3)$$
The action is invariant under a local scaling of the metric, or Weyl
transformation, which is parametrised by an infinitesimal function $\rho$,
$\delta_\rho\met=\rho \met$. For a critical string (D=26)
we can treat this as a
gauge symmetry. The action is also invariant under an infinitesimal
change of the world-sheet coordinates parametrised by the infinitesimal
vector $\zeta\sp a$, i.e. $\delta_\zeta
\met=\nabla_a\zeta_b+\nabla_b\zeta_a,\quad
\delta_\zeta X\sp \mu =\zeta\sp a\partial X\sp \mu /\partial \xi\sp a$. This is
also a
gauge symmetry. The conventional (i.e. Yang-Mills-like) Faddeev-Popov
treatment of the tree-level contribution to (3.1) was given in [10].
Using the general approach of the previous section we can consider the
contribution of an arbitrary Riemann surface. If we look for a variation
of the $\met$ that is orthogonal to the gauge variations $\delta_\rho\met$
and $\delta_\zeta\met$ then since the world-sheet is closed we require
that the variation satisfy $g\sp {ab}\delta\met =0$ and $\nabla\sp
a\delta\met=0$.
On a Riemann surface with $h>1,\quad h=1, $ and $h=0$
handles there are respectively $6h-6,2,$ and $0$ independent real
solutions to these equations, $\psi\sp A_{ab}$, and these zero-modes are in one
to
one
correspondence with the moduli which parametrise those changes of $\met$
that are not gauge-transformations. As a gauge-fixing condition
it is conventional to take the background gauge-condition $\met-\newhat \met
(t)=0$, where $\newhat \met$ is an arbitrary metric depending on the $m$
moduli,
$\{t\sp A\}$.Ultimately we require that the partition function, as well as
transition
amplitudes, be independent of this reference metric.
\medskip
     Carrying out the procedure of the previous section we introduce a
ghost, $c$, for each Weyl transformation,  $\rho$, a ghost, $c\sp a$, for each
reparametrisation  $\zeta\sp a$, anti-ghosts
and Lagrange multipliers $b\sp {ab},\lambda\sp {ab}$ corresponding to the
constraints, and quasi-ghosts $\tau\sp A$ for each modulus $t\sp A$.
The
Faddeev-Popov action is then

$$S_{FP}[\Psi]=S[\met ,X\sp \mu ]+i\int d\sp 2\xi\,\lambda\sp {ab}(\met-\newhat
\met )$$
$$+\int d\sp 2\xi (cb\sp {ab}\met +2\nabla_ac_b b\sp {ab}-\tau\sp A b\sp
{ab}\partial_A
\newhat \met). \eqno (3.4)$$
The $h$-string-loop contribution to the partition function $z(t)=\int\D\Psi\,
exp(-S_{FP})$ can be simplified by performing a number of integrations.
Firstly we integrate over the Lagrange multipliers to re-obtain
delta-functions for the constraints. These are used to perform the
integral over $\met$. The integral over the Weyl ghost may be performed
to produce another set of delta-functions, this time imposing the
tracelessness of the anti-ghosts, so these are used to integrate over the
trace of the anti-ghost. Finally, we integrate over the `quasi-ghosts',
$\tau$, to obtain $6h-6$ anti-ghost insertions. The result is [11]

$$z(t)=\int \D (X\sp \mu ,b\sp {ab},c\sp a)
\,\left(\prod_{A=1}\sp m \int d\sp 2\xi\,
b\sp {ab}\partial_A\newhat \met \right)\, e\sp {-S[\newhat \met ,X\sp \mu
]-\int d\sp 2\xi\,
2\nabla_a c_b \, b\sp {ab}},\eqno (3.5)$$
where $\newhat \met b\sp {ab}=0$.
If
we decompose the anti-ghost into a piece in the space of zero-modes
$\{\psi\sp A\}$ and an orthogonal piece as $b_{ab}=b\sp A\psi_{A\,ab}+\newhat
b_{ab}$,
then
the $b\sp A$ do not appear in the exponent but only in the insertions,
so integrating over them generates the determinant of $(\psi_A,\partial _B
\met )\equiv m_{AB}$. From (2.24) the effect of varying $\newhat \met$ on the
integral of $z(t)$ over the cut-off
moduli space is

$$\delta\int_M dt\,z(t)
=-\int_{\partial M}d\Sigma_A\int \D \Psi\,\tau\sp A
\left(\int d\sp 2\xi \,b\sp {ab}\delta\newhat \met\right)\,e\sp {-S_{FP}}.\eqno
(3.6)$$
Performing
the same integrations as before we arrive at

$$\delta\int_M dt\,z(t)=-\int_{\partial M}d\Sigma_A\int\D (X\sp \mu ,b\sp {ab},
c\sp a )
\,\tau\sp A\,(\int d\sp 2\xi\,b\sp {ab}\delta\newhat \met )$$
$$\times\left(\prod_{B\neq A} \int d\sp 2\xi\,
b\sp {ab}\partial_B\newhat \met \right)\, e\sp {-S[\newhat \met ,X\sp \mu
]-\int d\sp 2\xi\,
2\newhat \nabla_a c_b \, b\sp {ab}},\eqno (3.7)$$
where
$\newhat \nabla$ is the connection constructed from $\newhat \met$.
Again, if we decompose the anti-ghost as $b_{ab}=b_A\psi\sp A_{ab}+\newhat
b_{ab}$,
then the $b_A$ can only appear in the insertions, so these produce a factor
of

$$-(\int d\sp 2\xi\,\psi\sp {ab}_B\,\delta\newhat \met)\,m\sp {-1}_{AB}\,det
\,m
.\eqno (3.8)$$
Now we can
decompose an arbitrary variation of $\delta\newhat \met$ as $\rho\newhat \met+
\nabla_a\zeta_b+\nabla_b\zeta_a +\delta t\sp A\partial_A\newhat \met$.
Substituting this into (3.8) just picks out $\delta t\sp A\, det\, m$, so
(3.7) becomes

$$\delta\int_Mdt\,z(t)=\int_{\partial M}d\Sigma_A\,\delta t\sp A\,z(t).\eqno
(3.9)$$
This simply states that the only changes in the background metric
on which the partition function depends are those corresponding to changes
in the cut-off modulus, compare (1.3). It is the divergences in the integration
over
this modulus that we need to cancel.
Rather than considering just the
partition function it is necessary to address the more stringent problem
of regulating divergences in the Ward identities (2.28) for arbitrary operators
$V(\Psi)$.
In [12] we used a formulation of the BRST invariance that was rather
special to string theory to show that moduli space divergences lead to a
breakdown in the invariance, and we proposed that re-establishing this
invariance should be a constraint on any attempt to control these
divergences, for example by the use of counter-terms. With the more
general approach to gauge-fixing of the present paper, which is as
applicable to Strings as to Yang-Mills theory, we again obtain a
breakdown in the BRST invariance. We will now study the one-loop effect,
and go further than [12] by explicitly constructing a tree-level
counter-term to re-instate the BRST invariance.
The divergence we consider occurs as $\Im \tau\rightarrow\infty$ in (1.2),
and is due to the tachyon. This is normally considered a pathology of the
theory, and so is not usually addressed. There are other divergences in
the theory, even when supersymmetry is included, and they have been
treated in [22].

     The operators
of interest are integrals over the world-sheet of vertex operators, i.e.
$\int d\sp 2\xi \,\sqrt g\, V$. For $h=0,1$ there are ghost zero-modes due to
the
existence of conformal Killing vectors,
these are vectors which generate reparametrisations of the metric
equivalent to Weyl transformations. The integral over the ghosts will
vanish unless there are insertions  to saturate these zero-modes.
For $h=0$ there are six real conformal Killing vectors, for $h=1$ there
are $2$. For $h=0$ it is convenient to replace three of the vertex operator
integrals
by insertions of the form $\epsilon_{ab}c\sp ac\sp bV$, and for $h=1$ just one
replacement is necessary. This also corrects for the
over-counting of the gauge degrees of freedom. If we associate one power
of the coupling with each vertex operator, then tree-level scattering
amplitudes, which have no moduli, are given by [11]

$${\cal A}(\{V_i\})_0=
\kappa\sp {(n-6)}\int \D\Psi\left(
\prod_{i=1}\sp 3\epsilon_{ab}\,c\sp a\,c\sp b\,V_i\right)\,
\left(\prod_{j=4}\sp n\int d\sp 2\xi\,\sqrt g\,V_j \right)e\sp {-S_{FP}}$$
$$=\kappa\sp {(n-6)}\int \D (X\sp \mu ,b\sp {ab},c\sp a)
\,
e\sp {-S[\newhat \met ,X\sp \mu ]-\int d\sp 2\xi\,
2\newhat \nabla_a c_b \, b\sp {ab}}$$
$$\times \left(
\prod_{i=1}\sp 3\epsilon_{ab}\,c\sp a\,c\sp b\,V_i\right)\,
\left(\prod_{j=4}\sp n\int d\sp 2\xi\,\sqrt{\newhat \met}\,V_j \right)\eqno
(3.10)$$
whereas
one-string-loop amplitudes, which have two real moduli
corresponding to the complex modulus of (1.2), are
given by

$${\cal A}(\{V_i\})_1=
\kappa\sp {n}\int_F dt\,\int \D\Psi\epsilon_{ab}\,c\sp a\,c\sp b\,V_1\,
\left(\prod_{j=2}\sp n\int d\sp 2\xi\,\sqrt g\,V_j \right)e\sp {-S_{FP}}$$

$$=\kappa\sp {n}\int_F dt\,\int \D (X\sp \mu ,b\sp {ab},c\sp a)
\,
e\sp {-S[\newhat \met ,X\sp \mu ]-\int d\sp 2\xi\,
2\newhat \nabla_a c_b \, b\sp {ab}}$$
$$\times \left(\prod_{A=1}\sp 2\int d\sp 2\xi \,b\sp {ab}\partial_A\newhat
\met\right)
\left(
\newhat \epsilon_{ab}\,c\sp a\,c\sp b\,V_1\right)\,
\left(\prod_{j=2}\sp n\int d\sp 2\xi\,\sqrt{\newhat \met} V_j \right)\eqno
(3.11)$$
$\newhat \epsilon$ is the anti-symmetric tensor constructed using $\newhat
\met$.

\medskip
     Consider now the Ward identities (2.28) for one-string-loop.
As an illustration we take $V(\Psi )$ to
have the form
$V=\int d\sp 2\xi b\sp {rs}
\partial_rX\sp \mu\partial_sX\sp \nu l_{\mu\nu}e\sp {ik_0\cdot X}
(\prod_{i=1}\sp {n-1}\int d\sp 2\xi \, V_i)
\epsilon_{ab}c\sp ac\sp bV_n$. This contains the anti-ghost, since otherwise
the
expectation value of $\varsigma V$ will vanish, as we will soon see.
On higher genus surfaces $V$ would need the anti-ghost anyway, in order that
$\varsigma V$ have ghost number zero, on the torus we need it to have ghost
number two in order to saturate the ghost zero-modes due to the conformal
Killing vectors. We will take  $k_0\sp 2=-8$ and $\eta\sp {\mu\nu}l_{\mu\nu}=
k_0\sp \mu l_{\mu\nu}=0$, in which case $b\sp {rs}\partial_rX\sp
\mu\partial_sX\sp \nu
l_{\mu\nu}e\sp {ik_0\cdot X}$ is the vertex operator of a level two state, the
BRST transform of which is a spurious state representing a gauge degree of
freedom at level two. We will take the  $V_i$  to be tachyon vertex
operators, $V_i=e\sp {ik_i\cdot X},\quad k_i\sp 2=8$.
Putting this in (2.28) we obtain the Ward
identity

$$\langle\varsigma V(\Psi )\rangle_1=
\int_{\partial F}d\Sigma_A\,\int{\cal D}\Psi \,\tau\sp A\,V(\Psi )\,e\sp
{-S_{FP}}
=$$
$$\int_{\partial M}d\Sigma_A\,\int \D (X\sp \mu ,b\sp {ab},c\sp a)
\,
e\sp {-S[\newhat \met ,X\sp \mu ]-\int d\sp 2\xi\,
2\newhat \nabla_a c_b \, b\sp {ab}}\left(
\int d\sp 2\xi\, b\sp {rs}
\partial_rX\sp \mu\partial_sX\sp \nu l_{\mu\nu}e\sp {ik_0\cdot X}\right)$$
$$\times\left (\epsilon_{ab}\,c\sp a\,c\sp b\,V_n\right)\,
\left(\prod_{j=1}\sp {n-1}\int d\sp 2\xi\,\sqrt {\newhat g} V_i\right)
\epsilon_{AB}\int d\sp 2\xi\,b\sp {ab}\partial_B\newhat \met
\eqno (3.12)$$
Were this to vanish it would express the decoupling from a scattering
process of the level two gauge
degree of freedom represented by the spurious state.
To compute the derivatives $\partial_A\newhat \met$ we take coordinates
independent of the moduli, so we fix $\newhat \met$ by taking $ds\sp 2=|d\xi\sp
1+\tau
d\xi\sp 2|\sp 2/\Im \tau$ and $0\le \xi\sp a\le1$, with opposite sides of the
unit
square identified.
The cut-off boundary $\partial F$
is defined by taking $\Im \tau =T>>1$.
Decomposing the anti-ghost as before, $b_{ab}=b\sp A\psi_{A\,ab}+\newhat
b_{ab}$, we
see that the integral over the $b\sp A$ gives zero unless, as we stated, $V$
contains $b\sp {rs}$. In fact the integral over the $b\sp A$ yields a factor of
$(det\,m)\,(m\sp {-1})_{AB}\psi\sp {rs}_B$. So that we can rewrite
(3.12) as

$$\int_{\partial F}d\Sigma_A\,
\int{\cal D}\Psi \, e\sp {-S_{FP}}\,
\left(\int d\sp 2\xi\,(m\sp {-1})_{AB}\psi_B\sp {rs}
\partial_rX\sp \mu\partial_sX\sp \nu l_{\mu\nu}e\sp {ik_0\cdot X}\right)$$
$$\times
\left(\epsilon_{ab}\,c\sp a\,c\sp b\,V_n\right)\,
\left(\prod_{j=1}\sp {n-1}\int d\sp 2\xi\,\sqrt {\newhat g}\,V_i\right).
\eqno (3.13)$$
The functional integral is
given by a standard calculation [7]. It is usual to express the result using
different world-sheet coordinates to the above. Take complex coordinates
$z=\xi\sp 1+\tau\xi\sp 2$ with the domain
$0\le
\Im z\le\Im\tau,\quad -{1\over 2}\le \Re z\le{1\over 2},\quad
z_n=\tau$.
The vertex operators in (3.13) are all Weyl invariant by virtue of the
anomalous behaviour of $e\sp {ik\cdot  X}$ [3,13]. This is true also for the
level
two operator because if we set $\psi_{A\,rs}=\sqrt g \tilde\psi_{A\, rs}$
then $\tilde\psi_{A\, rs}$ is Weyl invariant. $m_{AB}$ is also Weyl
invariant. So, in terms of these coordinates we can write (3.13) as

$$\int_{\partial F}d\Sigma_A\,
\int{\cal D}\Psi \, e\sp {-S_{FP}}\,
\int dz\,d\bar z\,(m\sp {-1})_{AB}\left(
\psi_{B\, zz}
\bar\partial X\sp \mu\bar\partial X\sp \nu l_{\mu\nu}
+\psi_{B\, \bar z\bar z}
\partial X\sp \mu\partial X\sp \nu l_{\mu\nu} \right)
e\sp {ik_0\cdot X}                  $$
$$\times
\left(\,c\sp z\,c\sp {\bar z}\,V_n\right)\,
\left(\prod_{j=1}\sp {n-1}\int dz\,d\bar z\,V_i\right),
\eqno (3.14)$$
where $\partial=\partial/\partial z,\bar\partial=\partial/\partial\bar z$.
This may be evaluated as $(\Im \tau )\sp {-2}\, C(\tau )\, F(\tau )$
with
\vfill
\eject

$$F(\tau )=\pi\sp n\Im\tau\int\left(\prod_{i=0}\sp {n-1}dz_i\bar z_i\right)
\left(\prod_{i<j}(\chi
(z_i-z_j))\sp
{k_i\cdot k_j}\right)P(\{z_i\})$$

$$P=(m\sp {-1})_{AB}
\tilde\psi_{B\,\bar z\bar z }\sum_{ij}
k_i\sp \mu l_{\mu\nu}k_j\sp \nu
(\partial \,ln\,\chi (z_0-z_i))(\partial \,ln\,\chi (z_0-z_j))$$
$$+(m\sp {-1})_{AB}\tilde\psi_{B\, z z }\sum_{ij}
k_i\sp \mu l_{\mu\nu}k_j\sp \nu
(\partial \,ln\,\chi (z_0-z_i))\sp \dagger
(\partial \,ln\,\chi (z_0-z_j))\sp \dagger ,$$
$$\chi(z)=2\pi\,sin\,\pi z\,\prod_1\sp \infty{1-2e\sp {2n\pi i \tau}cos\,2\pi
z\,+
e\sp {4n\pi i\tau}
\over (1-e\sp {2n\pi i\tau})\sp 2}, \eqno (3.15)$$
On $\partial F \quad\Im\tau=T$ and we can approximate $\chi (z)\simeq
2\pi z$.
We now change the integration variables to $\xi_i=
exp\,-2\pi i(z_i-iT/2)$ with integration ranges
$exp-T/2<|\xi |<exp\,T/2$. Computing the Jacobian, and taking account of
the mass-shell conditions, and the conditions on the polarisation
$l_{\mu\nu}$ we obtain  the leading order contribution to (3.13)

$$2\left({T\over 2}\right)\sp {-13}\,e\sp {4\pi T}\pi\sp n
\int\left(\prod_1\sp {n-1}d\xi_i d\bar\xi_i\right)\left(\prod_{n>i>j}
|\xi_i-\xi_j|\sp {k_i\cdot k_j/2}\right)$$
$$\times (m\sp {-1})_{AB}(
\tilde\psi_{B\,\bar\xi\bar\xi}\sum_{ij}k_i\sp \mu l_{\mu\nu}k\sp \nu_j
(\xi_0-\xi_i)\sp {-1}(\xi_0-\xi_j)\sp {-1}$$
$$+\tilde\psi_{B\,\xi\xi}\sum_{ij}k_i\sp \mu l_{\mu\nu}k\sp \nu_j
\left((\xi_0-\xi_i)\sp {-1}(\xi_0-\xi_j)\sp {-1}\right)\sp \dagger ).\eqno
(3.16)$$
The
integral is almost a tree-level amplitude, ${\cal A}_0$, for
world-sheet metric $ds\sp 2=d\xi d\bar\xi$, and $\xi$ having the whole
complex plane as its domain, and vertex operators $V_0= (
\tilde\psi_{B\,\bar\xi\bar\xi}\partial X\sp \mu\partial X\sp \nu
+\tilde\psi_{B\,\xi\xi}\bar\partial X\sp \mu\bar\partial X\sp \nu ) l_{\mu\nu}
e\sp {ik_0\cdot X}$ and $V_i=e\sp {ik_i\cdot X}$.
Apart from the
fact that the position of only the n-th tachyon vertex has been fixed and
the integration range of the $\xi_I$ is not the whole complex plane. So
as $T\rightarrow\infty$ the integral approaches the tree-level amplitude
multiplied by a divergent function of $T$, which we will call $v(T)$
to take account of the gauge symmetry generated by the conformal Killing
vectors.
Thus

$$\langle\varsigma V(\Psi )\rangle_1\simeq
\left(\int_{\partial F}d\Sigma_A\,
(m\sp {-1})_{AB}\,
2\left({T/2}\right)\sp {-13}e\sp {4\pi T}v(T)\right)
{\cal A}(\{ V_i\} )_0 \eqno (3.17)$$
This may be compared to the tree-level Ward-identity. The corresponding
operator is $V=\int d\sp 2\xi b\sp {rs}
\partial_rX\sp \mu\partial_sX\sp \nu e\sp {ik_0\cdot X}
(\prod_{i=1}\sp {n-4}\int d\sp 2\xi\sqrt g \, V_i)
\prod_{j=n-3}\sp n\epsilon_{ab}c\sp ac\sp bV_j$.
Because there are no moduli the Ward identity is just

$$\langle \varsigma V\rangle _0=\int \D\Psi\,\varsigma V(\Psi)\,e\sp {-S_{FP}}
=0\eqno (3.18)$$
However,
since (3.17) factorises into a product of a
tree-level amplitude and an integral over $\partial F$ we can hope to
cancel the right-hand-side of (3.12) by adding a counter-term to the
tree-level contribution. To this end consider the tree-level Ward
identitiy for the product of $V$ and a counter-term $\Lambda (\Psi)$.
Taking into account the Grassmann character of $\varsigma$ and $V$ we have
$\langle\varsigma(V\Lambda)\rangle_0=\langle(\varsigma V\Lambda-V\varsigma
\Lambda)\rangle_0=0$. If we take $\Lambda=\int d\sp 2\xi\sqrt g
\,g\sp {rs}u_{rs}$ with $u$ a world-sheet tensor independent of $\Psi$, then
$\varsigma\Lambda=
\int d\sp 2\xi\,\sqrt g (\nabla\sp rc\sp s +\nabla\sp sc\sp r-g\sp
{rs}\met\nabla\sp ac\sp b)u_{rs}
$, so

$$\langle V\varsigma \Lambda\rangle_0=
\kappa\sp {(n-6)}\int \D (X\sp \mu ,b\sp {ab},c\sp a)
\,
e\sp {-S[\newhat \met ,X\sp \mu ]-\int d\sp 2\xi\,
2\newhat \nabla_a c_b \, b\sp {ab}}$$
$$\times \left(\int d\sp 2\xi\, b\sp {rs}
\partial_rX\sp \mu\partial_sX\sp \nu\,l_{\mu\nu}\, e\sp {ik_0\cdot
X}\right)\left(
\int d\sp 2\xi\,\sqrt {\newhat g} (\newhat \nabla\sp rc\sp s +\newhat \nabla\sp
sc\sp r-\newhat g\sp {rs}\newhat
\met\newhat \nabla\sp ac\sp b)u_{rs}\right) $$
$$\times
\left(\prod_{i=n-3}\sp n\newhat \epsilon_{ab}\,c\sp a\,c\sp b\,V_i\right)\,
\left(\prod_{j=1}\sp {n-4}\int d\sp 2\xi\,\sqrt{\newhat \met}\,V_j \right)\eqno
(3.20)$$
The contraction of $b\sp {rs}$ and the ghosts in $\varsigma\Lambda$
results in $u_{rs}$,  and the remainder of the expression is
just the tree-level amplitude ${\cal A}(\{V_i\})_0$.
If we take

$$u_{rs}=
\int_{\partial F}d\Sigma_A\,
(m\sp {-1})_{AB}\,\tilde\psi_{B\,rs}
2\left({T/2}\right)\sp {-13}e\sp {4\pi T}v(T), \eqno (3.21)$$
then
$\langle\varsigma V\rangle_1=\langle(\varsigma V)\Lambda\rangle_0$.
The full amplitude for n vertex operators is a sum
weighted by powers of the coupling, $\sum_{h=0}\sp \infty
\kappa\sp {(6h-6+n)}\langle V\rangle _h$.
We will now modify the tree-level contribution by including $\Lambda$ as a
counter-term, so we define a new sum to one-string-loop order

$$\langle\langle V\rangle\rangle=\kappa\sp {(n-6)} \langle V\,(1-
\kappa\sp 6\Lambda)\rangle_0+\kappa\sp n\langle V\rangle_1.\eqno (3.22)$$
This
is constructed so that $\langle\langle \varsigma V\rangle\rangle=0$, which
means that the new sum is independent of how we parametrise the functional
integration over $\Psi$. Note that the counter-term is equivalent to a
modification of the action $S_{FP}\rightarrow S_{FP}-\kappa\sp 6\Lambda$.

\bigskip
\bigskip
\centerline{\bf 4. Yang-Mills}
\bigskip
 We now turn to the problem of moduli space divergences in the
expansion of Yang-Mills theory as a sum over instanton sectors,
labelled by instanton number, $n$.
The partition function in a Euclidean space-time with metric $g_{\mu \nu}$
is formally [4]

$$Z=\sum_{n=-\infty}\sp \infty e\sp {-in\theta}\int_n\D A\,e\sp {-S_{YM}},$$

$$S_{YM}=-{1\over 4g\sp 2}\int d\sp 4 x\sqrt g g\sp {\mu\rho}g\sp {\nu\sigma}\,
tr\,({\bf F}_{\mu\nu}{\bf F}_{\rho\sigma}).\eqno (4.1)$$
The
field-strength is ${\bf F}_{\mu\nu}=[\partial_\mu+{\bf A}_\mu,
\partial_\nu+{\bf A}_\nu]$ and the gauge potential is an element of a Lie
algebra which we take to be su(N),
${\bf A}=A\sp aT_a, \quad T_a\sp \dagger=-T_a,\quad [T_a,T_b]=f_{ab}\sp cT_c,$
and
$tr\,(T_aT_b)=-\delta_{ab}$. The theta-angle plays the role of a coupling
for the expansion.
The action is invariant under local scalings of the metric, and so if we
take $g_{\mu\nu}=\delta_{\mu\nu}\Omega\sp 2 (x) $ then it is independent of
$\Omega$, consequently the classical equations of motion are conformally
covariant, which means that if we apply a conformal transformation to one
solution we obtain another.
However, the need to introduce a mass-scale in the quantisation
of the theory breaks this invariance and there is a Weyl anomaly. That is,
the regulated volume element $\D A$ does not share the invariance [14]. In the
one-instanton sector the action is minimised by the classical solution [15]

$$A_\mu\sp a=\eta_{\mu\nu}\sp a\,{2(x-y)\sp \mu\over (x-y)\sp 2+\rho\sp
2}\equiv
{\cal A}_1(x;y,\rho)\eqno (4.2)$$
$(x-y)\sp 2=(x\sp \mu -y\sp \mu )(x\sp \mu -y\sp \mu )$ and the
$\eta\sp a_{\mu\nu},\,a=1,2,3$ form a basis for self-dual tensors.
The parameters $y\sp \mu$ and $\rho$ are a complete set of moduli for this
sector.
With
the constraints (2.6) these yield 't Hooft's one-loop result (1.1) for the
one-instanton sector partition function when $\Omega=1$. As stated earlier
this diverges for large $y\sp \mu$ and $\rho$. This implies that
BRST invariance is broken as the Ward identities are replaced by (2.28).
This infra-red divergence is usually thought to be an artefact of the
approximation, which due
to asymptotic freedom is only really valid at short distances and scales.
However, it was pointed out in [16] that when space-time is compactified
to $S\sp 4$, as it should be to control the infra-red behaviour of the
one-loop calculation, the large
$y\sp \mu,\,\rho$ behaviour is reliably computable and not given by (1.1),
but rather, leads to a convergent integral. There is then no problem with
BRST invariance. We will outline this argument and then discuss the
problem of divergences in the two-loop sector which, in contrast to those
in the one-loop sector, include apparent short distance divergences.

\medskip
      For a spherical space-time of radius $a$ the function
$\Omega=2/(1+x\sp 2/a\sp 2)$, and the one-instanton partition function may be
obtained by integrating the Weyl anomaly for (1.1). However, it is not
necessary to do this to discover the limiting behaviour of the integrand.
For small $y\sp \mu$ and $\rho$ the integrand is approximately the same as in
the case of $R\sp 4$ because at short distances $R\sp 4$ and $S\sp 4$ look the
same.
Now there is an invariance of the metric corresponding to inversion
through the centre of the sphere followed by a parity transformation to
reinstate the original orientation. In terms of coordinates
$x\sp \mu\rightarrow \tilde x\sp \mu=a\sp 2\, m\sp \mu_\nu\, x\sp \nu/x\sp 2,\,
m=diag\,(1,-1,-1,-1)$ and $(1+\tilde x\sp 2/a\sp 2)\sp {-2}d\tilde x\sp 2=
(1+x\sp 2/a\sp 2)\sp {-2}dx\sp 2$. This is a conformal transformation so its
effect on the
instanton solution is equivalent to a change of the moduli, i.e.

$${\cal A}_\mu (\tilde x ;\rho,y)\,d\tilde x\sp \mu =
{\cal A}_\mu ( x ;\tilde\rho,\tilde y)\,d x\sp \mu ,\quad
\tilde\rho={a\sp 2\rho \over y\sp 2+\rho\sp 2},\quad \tilde y\sp \mu=
{a\sp 2 m\sp \mu_\nu y\sp \nu \over y\sp 2+\rho\sp 2}.\eqno (4.3)$$
The
partition function is reparametrisation invariant under changes in the
space-time  coordinates and so is the same whether we use $x\sp \mu$ or
$\tilde x\sp \mu$. Consequently
for small $\tilde y,\tilde\rho$ where the semi-classical approximation is
reliable

$$Z_1\sp {YM}=e\sp {-{8\pi\sp 2\over g\sp 2(\mu )}}\int
{d\sp 4\tilde y\, d\tilde\rho\over\tilde\rho\sp 5}\tilde\rho\sp {11N/3}.\eqno
(4.4)$$
Furthermore the measure $d\sp 4y\,d\rho /\rho\sp 5$ is
invariant under the action of conformal transformations on the moduli,
so $d\sp 4y\,d\rho /\rho\sp 5=d\sp 4\tilde y\,d\tilde\rho /\tilde\rho\sp 5$,
and

$$Z_1\sp {YM}=e\sp {-{8\pi\sp 2\over g\sp 2(\mu )}}\int
{d\sp 4y\, d\rho\over\rho\sp 5}\tilde\rho\sp {11N/3}
=e\sp {-{8\pi\sp 2\over g\sp 2(\mu )}}\int
{d\sp 4y\, d\rho\over\rho\sp 5}\left({a\sp 2\rho \over y\sp 2+\rho\sp
2}\right)\sp {11N/3},
\eqno (4.5)$$
which
is valid for small $\tilde\rho,\tilde y\sp \mu$ which means {\it large}
$\rho,y\sp \mu$. Hence the moduli space integration converges in the
one-instanton sector.
\medskip
        The two-instanton contribution to the partition function has
divergences that appear to be short distance effects. The Yang-Mills
action is minimised by the configuration [17]

$${\cal A}_2(x;y_0,y_1,y_2,\lambda_1,\lambda_2)={1\over 2}\eta\sp a_{\mu\nu}
{\partial\over\partial x\sp \nu}ln\,\phi ,\quad
\phi=\sum_{i=0}\sp 2 {\lambda_i\sp 2\over (x-y_i)\sp 2},\quad \lambda_0=1.\eqno
(4.6)$$
$\lambda_1,\lambda_2,y_0\sp \mu,y_1\sp \mu,y_2\sp \mu$ provide one too many
parameters to coordinatise the two instanton moduli space,
so there is a gauge symmetry amongst them.
When the space-time is taken to be $R\sp 4$ the partition function is [18]

$$Z_2\sp {YM}=e\sp {-{16\pi\sp 2\over g\sp 2(\mu )}}\left({4\pi\over g\sp
2(\mu)}\right)\sp 8
e\sp {-2\alpha(1)}{4\pi\over 3}\int {d\lambda_1\over\lambda_1}
{d\lambda_2\over\lambda_2}\,
d\sp 4y_0\,d\sp 4y_1\,d\sp 4y_2\,W\sp 4 \,{N_A\sp {4/3} \over N_S\sp
{1/3}}\sqrt\Gamma ,$$

$$W=z\sp 3\lambda_1\lambda_2,\quad z\sp 2={1\over 1+\lambda_1\sp 2+\lambda_2\sp
2},$$
$$N_A=z\sp 2(\lambda_2\sp 2(y_0-y_1)\sp 2+\lambda_0\sp 2(y_1-y_2)\sp 2
+\lambda_1\sp 2(y_2-y_0)\sp 2),$$

$$N_S=W\sp 2(y_0-y_1)\sp 2(y_1-y_2)\sp 2(y_2-y_0)\sp 2 ,$$
$$\Gamma=\Gamma ((y_0-y_1)\sp 2,\,(y_1-y_2)\sp 2,\,(y_2-y_0)\sp 2 ),$$
$$\Gamma (a,b,c)=2(ab+bc+ca)-a\sp 2-b\sp 2-c\sp 2.\eqno (4.7)$$
$\alpha$ is a function tabulated in [6].
The
$N_A\sp {4/3}/N_S\sp {1/3}$ contribution is approximate [19], but is accurate
near
configurations which degenerate into those of lower instanton number,
which is the region of interest to us. These are the configurations for
which $(y_i-y_j)\sp 2\rightarrow 0$. The integrand itself does not diverge
too badly near these configurations, for example as
$y_2\rightarrow y_1$ it behaves as $((y_1-y_2)\sp 2)\sp {-1/3}$. However it
then
depends
on $y_0$ and $y_1$ only through the difference $y_0-y_1$, so that if the
integrals over $y_1$ and $y_0$ are written as integrals over the
difference $y_0-y_1$ and the average $(y_0+y_1)/2$ then this last
integration gives an infinite volume factor. Thus the partition function
appears to diverge as the small distance  $(y_1-y_2)\sp 2$ goes to zero.
The obvious question to consider is whether the partition function for
$S\sp 4$ has this divergence. We will see that, as in the one instanton
sector, compactifying the theory on $S\sp 4$ removes these moduli space
divergences.
\medskip
     The metric on $S\sp 4,\,
g_{\mu\nu}=(2/(1+x\sp 2/a\sp 2))\sp 2\delta_{\mu\nu}=\Omega\sp
2\delta_{\mu\nu}$
differs from that on $R\sp 4,\, g_{\mu\nu}=\delta_{\mu\nu}$ by a local scaling.
So, to construct the partition function on $S\sp 4$ we study the effect of a
Weyl transformation on the metric. The classical action, $S_{YM}$, is Weyl
invariant, but the regulated volume element $\D A$ is not, so there is a
Weyl anomaly. The change in the partition function resulting from a
symmetry of the classical action that commutes with the gauge invariance,
and is respected by the volume element, was given by (2.26). This must now
be modified to take account of the anomaly. Thus, under the transformation
$\delta_\phi g_{\mu\nu}=\phi g_{\mu\nu}$ the volume element changes, so

$$\delta_\phi\int_M dt\,z(t)=\int_Mdt\,\int \D\Psi\left(\int d\sp 4x\,\phi
W(x)\right)e\sp {-S_{FP}}-\int_{\partial M}d\Sigma_A\int\D
\Psi\tau\sp Ab\sp j\left(\delta_\phi  F_j\right)\,e\sp {-S_{FP}}.\eqno (4.8)$$
If
we work to one-loop the integral over the cut-off boundary does not
contribute with the choice of $F_j$ (2.6) which are linear in the `quantum
correction,' ${\bf A}-{\cal A}$, and so contribute at higher order in the
expansion in powers of Planck's constant.
If $z_a(t)$ and
$\D _a\Psi$ denote  the
partition function moduli space density and
volume element for a sphere of radius $a$, then
taking $\phi=\delta a{d\over da}ln\,\Omega\sp 2$ gives, to one-loop

$${d\over da}\int_Mdt\,z_a(t)=\int_M dt\,\int\D_a\Psi\left(\int d\sp 4 x\,
{d\,ln\,\Omega\sp 2\over da}
W\right)\,e\sp {-S_{FP}}. \eqno (4.9)$$
Integrating with respect to $a$ from $\infty $ to $a$ we obtain

$$\int_Mdt\,z_a(t)=\int_M dt\,\int\D_\infty\Psi\,
exp\left(-S_{FP} +\int d\sp 4 x\,ln\,(\Omega\sp 2/4)\,W\right). \eqno (4.10)$$
On general grounds [20] the anomaly
density $W$ is a sum of the Lagrangian and the Euler density for the
sphere [21]. The latter makes a
contribution independent of the moduli which we will ignore. To one loop
we  evaluate $W$ for the background field ${\cal A}$, so to this order

$$\int d\sp 4x \,ln\,(\Omega\sp 2)\,W={\beta (g\sp 2) \over 4 g\sp 4}\int d\sp
4x\,
\,ln (\Omega\sp 2)\, tr\, (F_{\mu\nu}F_{\mu\nu}).\eqno (4.11)$$
Where the beta-function is

$$\beta =\mu{\partial(g\sp 2(\mu))\over\partial\mu}=-{g\sp 411N\over 24\pi\sp
2}.
\eqno (4.11)$$
For
the instanton solution (4.6) we have [17]
$$tr\, F_{\mu\nu}F_{\mu\nu}=
2\partial\sp 2\partial\sp 2\,ln\, \sigma ,$$
$$ \sigma\equiv
\left(\lambda_0\sp 2(x-y_1)\sp 2(x-y_2)\sp 2
+\lambda_1\sp 2(x-y_2)\sp 2(x-y_0)\sp 2+\lambda_2\sp 2(x-y_0)\sp 2(x-y_1)\sp 2
\right),       \eqno (4.12)$$
where   $\partial\sp 2$ is the flat four-dimensional Laplacian.
When
$y_2=y_1$ the function $\sigma$ has a factor of $(x-y_1)\sp 2$, but
$\partial\sp 2\partial\sp 2 \,ln\,(x-y_1)\sp 2=-16\pi\sp 2\delta (x-y_1)$,
so that $exp\,\int d\sp 4x\,ln(\Omega\sp 2/4)\,W$ depends on $y_1$ via the
factor
$(1+y_1\sp 2/a\sp 2)\sp {-22N/3}$. The presence of this in the partition
function
density $z_a(t)$  makes the integral over $(y_0+y_1)/2$ converge. The same
arguments may be applied to the divergences of the partition function on
$R\sp 4$ as the scales $\lambda_i\rightarrow 0$. Thus the moduli space
divergences are indeed removed by compactifying the theory on $S\sp 4$.
\bigskip
\bigskip
\centerline{\bf 5. Conclusions}
\bigskip
    The configuration spaces of many field theories consist of
disconnected pieces. Quantum mechanical amplitudes are then sums over
these components weighted by a function of a coupling constant. For
each component the amplitudes can be reduced to finite dimensional
integrals over moduli. Within the semi-classical approach
the moduli appear as the parameters that label physically
distinct solutions to the equations of motion. In String Theory however,
they label surfaces that are not equivalent under
reparametrisations and local scalings of the metric.
We used the Faddeev-Popov trick to make the dependence on the moduli
explicit without resorting to approximation.
This involves choosing coordinates on the configuration space by imposing
constraints.
In general the moduli space
integrals will diverge. If we regulate them with cut-offs the integrals
will depend strongly on the choice of cut-off surfaces in moduli space.
As a result amplitudes are not independent of the constraints
we use to parametrise configuration space. This dependence is expressed by
(2.26) and is the origin of the `BRST anomaly' (2.28). Physics must be
independent of such arbitrary choices, and so a viable quantum theory must
be capable of being cured of this disease. The two theories
we have considered in detail,
Bosonic Strings and Yang-Mills, cope with this in different ways. In String
Theory the one-string-loop divergence is due to a toroidal world-sheet
degenerating by becoming infinitely long. Scattering amplitudes computed
for such a surface are given by the corresponding amplitudes for a
spherical world-sheet multiplied by a divergent factor. Thus it is
possible to cancel the one-loop `BRST anomaly' by  a tree-level
counter-term (3.22). It is common that moduli space divergences are
associated with configurations in a given component degenerating to those
of another component. This gives the possibility of `topological
renormalisation' in which the pathologies of the topological expansion
over the disconnected components of configuration space can be cured by
counter-terms relating different components. In Yang-Mills theory the
story is quite different. The
divergences associated with the integration over instanton moduli in the
one instanton sector disappear when the infra-red divergences of
the theory are properly regulated by compactifying onto $S\sp 4$
[16]. We have shown that the same thing happens to the divergences of the
two-instanton sector. Although we have worked with theories that have a
local invariance that requires gauge fixing, our considerations are not
restricted to such theories.

\vfill
\eject

\centerline{\bf References}
\bigskip

$$\vbox{\halign{#\hfil & \hskip 0.1in#\hfil\cr [1]& L.D.Faddeev, V.N.Popov
Phys.Lett. B25(1967)29\cr
[2]& C.Becchi, A.Rouet, R.Stora Phys.Lett. B52(1974)344\cr
[3]& A.M.Polyakov Phys.Lett. B103(1981)207\cr
[4]& M.B.Halpern, S.A.Klein, J.A.Shapiro Phys.Rev. 188(1969)2378\cr
   & K.Kikkawa, S.Klein, B.Sakita, M.A.Virasoro Phys.Rev.D1(1970)3258\cr
   & D.J.Gross, A.Neveu, J.Scherk, J.H.Schwarz Phys.Lett. B31(1970)592\cr
[5]& R.Jackiw Rev.Mod.Phys. 52(1980)4\cr
[6]& G.'t Hooft Phys.Rev. D14(1976)3432\cr
[7]& J.A.Shapiro Phys.Rev D5(1972)1945\cr
[8]& G.'t Hooft, M.T.Veltman Nucl.Phys. B50(1972)318\cr
   & A.A.Slavnov Theor. and Math.Phys. 10(1972)99\cr
   & J.C.Taylor Nucl.Phys. B33(1971)436\cr
[9]& L.Brink, P.Di Vecchia, P.Howe Phys.Lett. B65(1976)471\cr
[10]& L.Beaulieu, C.Becchi, R.Stora Phys.Lett. B180(1986)55\cr
[11]& D.Friedan, E.Martinec, S.Shenker Nucl.Phys. B271(1986)93\cr
[12]& E.Cohen, C.Gomez, P.Mansfield Phys.Lett. B174(1986)159\cr
    & P.Mansfield Nucl.Phys. B283(1987)551\cr
[13]& E.D'Hoker, D.H.Phong Rev.Mod.Phys. 60(1988)917\cr
[14]& K.Fujikawa Phys.Rev D29(1984)285\cr
[15]& A.A.Belavin, A.M.Polyakov, A.S.Schwartz, Y.S.Tyupkin Phys.Lett
B(1975)85\cr
[16]& A.Patrasciouiu, A.Rouet Phys.Lett. B112(1982)472\cr
[17]& R.Jackiw, C.Nohl, C.Rebbi Phys.Rev. D15(1977)1642\cr
[18]& P.Goddard, P.Mansfield, H.Osborn Phys.Lett. B98(1981)59\cr
    & P.Mansfield Nucl.Phys. B186(1981)287\cr
    & H.Osborn Ann.Phys. 135(1981)373\cr
[19]& H.Osborn, G.Moody Nucl.Phys. B173(1980)422\cr
[20]& L.Bonora, P.Cotta-Ramusino, C.Reina Phys.Lett. B126(1983)305\cr
[21]& E.Corrigan, P.Goddard, H.Osborn, S.Templeton Nucl.Phys.
B159(1979)469\cr
[22]& E.Martinec Nucl.Phys. B281(1987)157\cr
    & C.G.Callan, C.Lovelace, C.R.Nappi, S.A.Yost Nucl.Phys.
B288(1987)525,\cr
    & B293(1987)83; Phys.Lett. B206(1988)41\cr
    & S-J Rey Nucl.Phys. B316(1989)197\cr
    & R.R.Metsaev, A.A.Tseytlin Nucl.Phys. B281(1987)157\cr}}$$

\vfill
\end
\bye